\documentclass[aps,pra,showpacs,twocolumn,superscriptaddress]{revtex4}

\usepackage{graphicx}
\usepackage{dcolumn}
\usepackage{bm}
\usepackage{amsfonts}
\usepackage{amsmath}
\usepackage{color}

\newcommand{\bra}[1]{\left\langle{#1}\right\vert}
\newcommand{\ket}[1]{\left\vert{#1}\right\rangle}
\newcommand{\beq}{\begin{equation}}
\newcommand{\eeq}{\end{equation}}
\newcommand{\bea}[1]{\begin{equation}\begin{array}{#1}}
\newcommand{\eea}{\end{array}\end{equation}}
\newcommand{\beqn}{\begin{eqnarray}}
\newcommand{\eeqn}{\end{eqnarray}}
\newcommand{\pp}{{\cal P}^2}
\newcommand{\cc}{{\cal C}^2}
\newcommand{\vv}{{\cal V}^2}
\newcommand{\conc}{{\cal C}}
\newcommand{\visi}{{\cal V}}
\newcommand{\prev}{{\cal P}}
\DeclareMathOperator{\tr}{Tr}  
\providecommand{\openone}{\mathbbm{1}}

\begin{document}


\title{Experimental investigation of the dynamics of entanglement: Sudden death, complementarity, and continuous monitoring of the environment}
\author{A. Salles}
\email{salles@if.ufrj.br}
\affiliation{Instituto de F\'{\i}sica, Universidade Federal do Rio de
Janeiro, Caixa Postal 68528, Rio de Janeiro, RJ 21941-972, Brazil}
\author{F. Melo}
\affiliation{Instituto de F\'{\i}sica, Universidade Federal do Rio de
Janeiro, Caixa Postal 68528, Rio de Janeiro, RJ 21941-972, Brazil}
\affiliation{Albert-Ludwigs-Universit\"at Freiburg, Physikalisches Institut, Hermann-Herder-Strasse 3, D-79104 Freiburg, Germany}
\author{M. P. Almeida}
\affiliation{Instituto de F\'{\i}sica, Universidade Federal do Rio
de Janeiro, Caixa Postal 68528, Rio de Janeiro, RJ 21941-972,
Brazil}
\affiliation{Centre for Quantum Computer Technology, Department of Physics, University of Queensland, QLD 4072, Brisbane, Australia.}
\author{M. Hor-Meyll}
\affiliation{Instituto de F\'{\i}sica, Universidade Federal do Rio
de Janeiro, Caixa Postal 68528, Rio de Janeiro, RJ 21941-972,
Brazil}
\author{S. P. Walborn}
\affiliation{Instituto de F\'{\i}sica, Universidade Federal do Rio
de Janeiro, Caixa Postal 68528, Rio de Janeiro, RJ 21941-972,
Brazil}
\author{P. H. Souto Ribeiro}
\affiliation{Instituto de F\'{\i}sica, Universidade Federal do Rio
de Janeiro, Caixa Postal 68528, Rio de Janeiro, RJ 21941-972,
Brazil}
\author{L. Davidovich}
\affiliation{Instituto de F\'{\i}sica, Universidade Federal do Rio de
Janeiro, Caixa Postal 68528, Rio de Janeiro, RJ 21941-972, Brazil}

%
%
\begin{abstract}
We report on an experimental investigation of  the dynamics of  entanglement between a single qubit and its environment, as well as for pairs of qubits interacting independently with individual environments, using photons obtained from parametric down-conversion.  The qubits are encoded in the polarizations of single photons, while the interaction with the environment is implemented by coupling the polarization of each photon with its momentum. A convenient Sagnac interferometer allows for the implementation of several decoherence channels and for the continuous monitoring of the environment. For an initially-entangled photon pair, one observes the vanishing of entanglement before coherence disappears. For a single qubit interacting with an environment, the dynamics of complementarity relations connecting single-qubit properties and its entanglement with the environment is experimentally determined.    The evolution of a single qubit under continuous monitoring of the environment is investigated, demonstrating that a qubit may decay even when the environment is found in the unexcited state.  This implies that  entanglement can be increased by local continuous monitoring, which is equivalent to entanglement distillation. We also present a detailed analysis of the transfer of entanglement from the two-qubit system to the two corresponding environments, between which entanglement may suddenly appear, and show instances for which no entanglement is created between dephasing environments, nor between each of them and the corresponding qubit: the initial two-qubit entanglement gets transformed into legitimate multiqubit entanglement of the Greenberger-Horne-Zeilinger (GHZ) type. 
\end{abstract}


\maketitle


%
%

\section{Introduction}
\label{sec:introduction}

Entanglement plays a central role in quantum mechanics. The
subtleties of  this phenomenon were first brought to light by the
seminal paper of Einstein, Podolski, and Rosen~\cite{EPR}, published
in 1935, and by those of
Schr\"odinger~\cite{schrodinger1,schrodinger2},  published in 1935
and 1936. It took however approximately thirty years for its
essential distinction from classical physics to be unmasked by John
Bell~\cite{bell}, and another thirty years for the discovery that
entanglement is a powerful resource for quantum
communication~\cite{ekert,bennett3,bouwmeester,boschi,gisin,bennett4}.
It was also found in the
90's to play an important role in quantum computation
algorithms~\cite{nielsen}. Furthermore, it plays a key role in the
behavior of macroscopic quantities like the magnetic susceptibility
at low temperatures~\cite{susceptibility}.

Yet the dynamics of entangled systems under the unavoidable effect
of the environment is still a largely unknown subject, in spite of
its fundamental importance in the understanding of the
quantum-classical transition, and its practical relevance for the
realization of quantum computers.

The absence of coherent superpositions of classically distinct
states of a macroscopic object is analyzed by decoherence
theory~\cite{joos,zurek:715}, which shows that the emergence of the
classical world is intimately related to the extremely small
decoherence time scale for macroscopic objects. Within a very short
time, which decreases with the size of the system, an initial
coherent superposition of two classically distinct states gets
transformed into a mixture, due to the entanglement of the system
with the environment. The decay dynamics is ruled, within a very
good approximation, by an exponential law.

Detailed consideration of the dynamics of entangled states requires defining proper measures of this quantity. For
pure states, one can use the Von Neumann entropy~\cite{nielsen}
associated to each part, or alternatively the corresponding purity,
defined by the so-called linear entropy~\cite{rungta01, mintert}. 
The ideais that the more entangled some partition of a
multiqubit state is, the more unknown is the state of each
part.

However, systems undergoing decoherence do not remain pure. A mixed
state of $N$ parties is separable if it can be written as a convex
sum of products of density matrices corresponding to each
part~\cite{werner}:
\begin{equation}
\label{Rho}
\varrho=\sum_{\mu} p_{\mu} \varrho_{1_{\mu}}\otimes\hdots\otimes\varrho_{N_{\mu}} ,
\end{equation}
where the index  $\mu$ refers to the $\mu$-th realization of the state and $\sum_{\mu} p_{\mu}=1$, with $p_{\mu}\geq0$.

Entanglement measures for mixed states have been defined for systems
with dimension up to six~\cite{wootters,peres,horodecki}, but for
larger dimensions this problem has not yet been solved. For
two-qubit systems, Wootters~\cite{wootters} introduced the
concurrence as a measure of entanglement.

It was shown by Peres~\cite{peres}  that,  if the partial transpose
of the density matrix of a multipartite system with respect to one
of its parts has negative eigenvalues, then the state is necessarily
entangled. Thus, a non-negative partial transpose is a necessary
condition for a state to be separable. However, this condition is
also sufficient only for $2\times2$ or $2\times3$ systems, as shown
in Ref.~\cite{horodecki}.

The negativity, defined as the magnitude of the sum of negative
eigenvalues of the partially transposed matrix, can thus be used, in
these cases, as a measure of entanglement. For higher dimensions,
this does not work anymore:   the negativity is then an indicator of
{\it distillable entanglement}. That is, if the negativity is
different from zero then it is possible, through local operations 
and classical communication, to obtain from $n$ copies of
the state a number $m$ ($m\le n$) of maximally entangled states. This process,
called distillation, does not work if the partially-transposed
density matrix is non-negative: any entanglement still present
cannot be distilled -- it is then called bound
entanglement~\cite{bound}.

These measures allow one to study the dynamics of initially entangled states under the influence of the environment~ \cite{mintert, karol, simon, diosi, dodd, dur, yu1, carvalho:230501, hein, fine:153105, santos:040305, yu:140403, almeida07, carvalho-2007, kimble07, aolita}. The outcome of these investigations is that the dynamics of entanglement can be quite different from the dynamics of decoherence: the first is not ruled by an exponential decay law, , as the latter, and entanglement can disappear at finite times, even when 
system coherences 
decay asymptotically in time. This phenomenon, known as \emph{entanglement sudden death}~\cite{yu:140403}, is a peculiar feature of global dynamics.

In this article we present an all-optical device to study the
interaction of simple systems (one or two qubits) with various kinds
of environments, in a highly controllable fashion. The setup is
extremely versatile, allowing to implement many different types of
open system dynamics. A partial account of our experimental results
was given in Ref.~\cite{almeida07}. Here we show how this set up can
be used not only to demonstrate the subtle dynamics of entanglement,
but also  the behavior of a
continuously-monitored system, as well as the dynamics of complementarity relations~\cite{englert, englert2, jakob-2003, melo} between local and
global properties for a two-qubit entangled system. These complementarity relations quantify
the notion that, for pure entangled states, coherences and populations of
each party become uncertain: the more unknown they are, the more entangled is
the state. 

Sections~\ref{sec:theory} and \ref{complementarity} contain the theoretical framework, in a form which is particularly suitable to the experimental investigation of the dynamics of entanglement under
different kinds of environment.  

Section~\ref{sec:theory} deals with open-system dynamics and Kraus operators, while Section~\ref{complementarity}
discusses quantum channels, the dynamics of complementarity for each of these channels, and the transfer of entanglement from the two-qubit system to the two corresponding environments.  A peculiar feature of dephasing processes is emphasized, for a family of initial states: when the two-qubit entanglement disappears, no bipartite entanglement is left in the system. The state of the two-qubit system plus corresponding environments becomes a state of GHZ (Greenberger-Horne-Zeilinger) type~\cite{GHZ}, with only genuine multiparticle entanglement. 

The experimental setup is introduced 
in section~\ref{sec:experiment}, 
along with several examples of environments that we are able to
implement. In section~\ref{sec:results_qubit} we present the experimental
results for the behavior of a single qubit, with and without continuous monitoring of the environment, including a detailed study of the dynamical behavior of complementarity relations between local (single-party) and global properties of the system qubit$+$environment. We show that our results on the continuously-monitored system are intimately related to the distillation of entanglement. 

In Section~\ref{sec:results_entanglement} we discuss the experimental investigation of the evolution of entanglement for two typical noise channels -- amplitude damping and dephasing -- including the first observation of the phenomenon of entanglement sudden death, which we
had previously reported in~\cite{almeida07}.  Our conclusions are summarized  in section~\ref{sec:conclusions}.

%
%

\section{Open system dynamics and Kraus operators}
\label{sec:theory}

A system ($S$) interacting with an environment ($E$) is described by
the following Hamiltonian: \beq H = H_S\otimes\openone
+\openone\otimes H_E + \lambda V_{SE}, \label{toth} \eeq where $H_S$
and $H_E$ are  the system and environment Hamiltonians respectively,
and  $V_{SE}$  is the coupling term between them with coupling
constant $\lambda$ (in the weak coupling limit, $\lambda \ll 1$).
The system and the environment get entangled due to the interaction
$V_{SE}$ -- an
initially  pure  state of $S$ evolves to a mixed state. 

In quantum optics, the traditional way of dealing with open systems
weakly coupled to environments with large number of degrees of freedom is through master
equations~\cite{lindblad, kossakowski}.  In this approach, the
equation of motion for the state $\rho_S$ of the system, given by:
 \beq
\dot{\rho}_S=-\frac{i}{\hbar}\tr_E [H,\rho_{SE}], \label{eqmotion}
 \eeq
where $\rho_{SE}$ is $S+E$ density matrix, is approximated to first order of perturbation theory, with
additional assumptions of Markov dynamics and initially uncorrelated
systems. The previous  expression can be then written as a sum of a
unitary contribution plus a non-unitary term, which depends only on
operators acting on the system $S$, and is given by the following
expression: 
\beq \dot\rho_S^{\rm NU} =
 - \sum_k \left(\rho_S{\cal
L}^\dagger_k {\cal L}_k +{\cal L}^\dagger_k {\cal L}_k \rho_S - 2
{\cal L}_k \rho_S  {\cal L}^\dagger_k\right)\,, \eeq 
where the upper
index NU stands for non-unitary, and $ {\cal L}_k $ are the
so-called Lindblad operators. See
References~\cite{carmichael, breuer} for a comprehensive treatment.

The experimental investigation of open system dynamics can be greatly simplified by adopting an alternative formalism, based on the Kraus representation~\cite{kraus}. We summarize in the following the main ingredients of this approach.

\subsection{Kraus operators}

\begin{figure}[b]
\includegraphics[width=7cm]{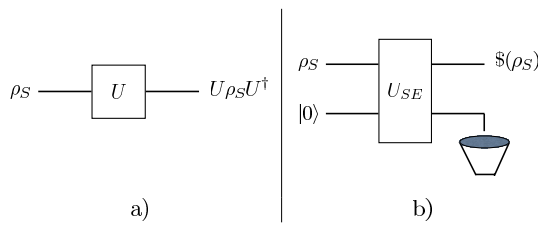}
\caption{\footnotesize Unitary dynamics. a) Closed system b) Open system --  $\$$ describes the reduced evolution of $S$ when we trace out the environment $E$.}
\label{openclose}
\end{figure}

As suggested in Eq.~(\ref{eqmotion}), the evolution of a system  coupled to an environment can always be expressed as a unitary dynamics on a higher dimensional system -- Fig.~\ref{openclose} depicts this approach.

Starting with uncorrelated systems, the total evolution can be written as:
\beq
U_{SE}(\rho_S\otimes\ket{0}_E\bra{0})U^\dag_{SE}\;;
\eeq
where $U_{SE}$ is the $S+E$ evolution operator and $\ket{0}_E$, without loss of generality,  represents the initial state of the environment. If we wish to focus only on the evolution of system $S$, we take the trace over the degrees of freedom of the environment. The effective evolution, not necessarily unitary, is then given by:
\beq
\$(\rho_S)\begin{array}[t]{l}
=\tr_E[U_{SE}(\rho_S\otimes\ket{0}_E\!\bra{0})U^\dag_{SE}]\;\\
=\sum_\mu\, \!_E\bra{\mu}U_{SE}\ket{0}_E\rho_S\,_E\!\bra{0}U^\dag_{SE}\ket{\mu}_E\;;
\end{array}
\label{TrKraus} 
\eeq 
where $\{\ket{\mu}\}$ form an orthonormal basis
for $E$, and the operator $\$$ describes the evolution of the system
$S$ ( $\$$ is usually called a quantum channel, in analogy with
classical communication theory~\cite{nielsen} ). Finally this
evolution can be expressed only in terms of operators acting on $S$
in the following form: 
\beq \$(\rho_S)=\sum_\mu M_\mu \rho_S
M_\mu^\dag, \label{EvolKraus}
 \eeq 
 where the operators 
 \beq\label{kraus}
 M_\mu\equiv\,
\!_E\bra{\mu}U_{SE}\ket{0}_E
\eeq
 are the so-called Kraus
operators~\cite{kraus,choi,preskill}. The property $\sum_\mu
M_\mu^\dag M_\mu = \openone$ guarantees that $\tr [\$(\rho_S)] =1$,
so that the operation $\$$ is trace preserving.  Furthermore, the
evolution given by Eq. (\ref{EvolKraus}) preserves the positive semi-definite character of $\rho_S$ -- this means that $\$(\rho_S)$ is also a
density operator.  It is important to note that the Kraus operators
are not uniquely defined -- performing the trace operation in
Eq.(\ref{TrKraus}) in different bases leads to  different sets of
equivalent operators,  yielding different decompositions of the
resulting density matrix. 

There are at most $d^2$ independent  Kraus operators~\cite{nielsen,leung:528}, where $d$ is the dimension of $S$. Together with Eq.~(\ref{kraus}), this property implies that, if $\{\ket{\phi_i}\}$ is a basis in the space corresponding to $S$, then a dynamical evolution of $S$, corresponding to the Kraus operators $\{M_\mu\}$, $\mu=0,\dots,d^2-1$, can be derived from a unitary evolution of $S+E$ given by the following map:
 \beq
\begin{array}{ccc}
\ket{\phi_1}\ket{0}&\rightarrow&M_0\ket{\phi_1}\ket{0}+ \dots + M_{d^2-1}\ket{\phi_1}\ket{d^2-1}\;;\\
\ket{\phi_2}\ket{0}&\rightarrow&M_0\ket{\phi_2}\ket{0}+ \dots + M_{d^2-1}\ket{\phi_2}\ket{d^2-1}\;;\\
\vdots&\rightarrow&\vdots\\
\ket{\phi_d}\ket{0}&\rightarrow&M_0\ket{\phi_d}\ket{0}+ \dots + M_{d^2-1}\ket{\phi_d}\ket{d^2-1}\;,
\end{array}
\label{KrausMaps}
\end{equation}
where as before the operators $M_i$ act only on $S$. This  map yields the guiding equations for our experiments. 

If the environment has many degrees of freedom (so that it can be considered a reservoir), then under Markovian and differentiability assumptions Eq.~(\ref{EvolKraus}) yields a master equation~\cite{preskill}. This is however less general than the  Kraus approach, which applies even if the environment has a small number of degrees of freedom. 

\subsubsection{Global vs. Local environments}

If the system $S$ is  itself composed of $N$ subsystems ($S_1\;,\;\dots\;,\;S_N$), we must distinguish between two main types of environment:

\noindent {\it i)} Global channels: in this case all the subsystems are embedded in the same environment, and can even communicate through it. These channels perform non-local dynamics and, in principle, can increase the entanglement among the subsystems.

\noindent {\it ii)} Local channels: each subsystem interacts with its own environment, no communication is present. The total evolution can be written as $U_{S_1E_1}\otimes\cdots\otimes U_{S_NE_N}$, and  Eq.(\ref{EvolKraus}) is replaced by:
\beq
\$(\rho_S)=\sum_{\mu\dots\nu} M^1_\mu\otimes\cdots\otimes M^N_\nu \rho_S{M^1_\mu}^\dagger\otimes\cdots\otimes {M^N_\nu}^\dagger\;.
\label{EvolKrausN}
\eeq
This operation is clearly local, and therefore cannot increase the entanglement among the constituents.

Obviously, for systems with $N>2$ mixed dynamics is also possible, i.e, some subsystems interact with a common environment and others with independent environments.

\subsubsection{Filtering operations -- Monitoring the environment}
\label{sec:environmentMonitoringTheory}

\begin{figure}[t]
\begin{center}
\includegraphics[width=6cm]{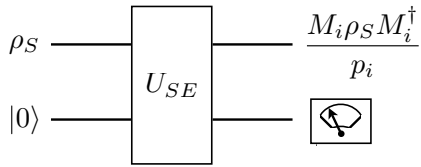}
\end{center}
\caption{\footnotesize Monitoring the environment. The state of the environment is measured, post-selecting the state of $S$.}
\label{filtering}
\end{figure}

Instead of directly following the dynamics of system $S$, one can infer it by monitoring its surroundings. For instance, by detecting a photon emitted by a two-level atom, we know for sure that the atom is in its ground state. This scheme is illustrated in Fig.~\ref{filtering}.

The formalism of the preceding sections must be changed to take into account the monitoring of the environment. Rather than tracing over the environment,  we perform a measurement on it. If the outcome $i$ is obtained, the state of $S$ evolves to:
\beq
\frac{M_i\rho_S M^\dagger_i}{p_i},
\eeq
where $p_i =\tr [M_i\rho_S M^\dagger_i]$ is the probability of finding the outcome $i$. Notice that, if the state $\rho_S$ is initially pure, it will remain pure after the  measurement on the environment.

This application of a single Kraus operator to the state is usually
called  a filtering  operation~\cite{verstraete:01, nielsen}. A
sequence of successive evolutions and measurements defines a quantum
trajectory for the state of $S$ -- each record of the state of the environment
defining a quantum jump~\cite{carmichael}.

%
%

\section{Complementarity, quantum channels, and the dynamics of entanglement}\label{complementarity}

Up to this point we were dealing with the open dynamics in a rather general way. From now on we  specialize  on systems composed of qubits, which are representative of many physical systems of interest for quantum information processing. Furthermore, we  consider only local environments, which implies in the interaction of each qubit only with its own environment. This is the situation for two decaying atoms separated by a distance much larger than the wavelength of the emitted radiation.  The individual qubit dynamics is then used to describe how the initial entanglement of two qubits is degraded due to the action of these independent environments.

An elegant way  to illustrate how the entanglement of the system
with the environment disturbs the individual properties of the
subsystems is through the complementarity  relations presented in
Ref.~\cite{jakob-2003}. This is described in the following
sub-section.

\subsection{Complementarity relations}
\label{sec:complementarity}

A single qubit $S$ in a pure state has two complementary aspects,
particle-like, and wave-like~\cite{bohr}, which  is mathematically
expressed by the following relation~\cite{englert,englert2}: \beq
\pp_S+\vv_S =1, \eeq where the ingredients are the single-particle
predictability $\prev_S$ and visibility $\visi_S$. The first is a
measure of the single-qubit relative population, defined as
$\prev_S = |\langle \sigma_z \rangle|$. The second is a measure of
single-qubit coherence and is defined as $\visi_S = 2 |\langle
\sigma^+ \rangle|$. Here $\sigma_i$, with $i \in\{x\,,y\,,z\}$, are
the Pauli matrices, and $\sigma^+=\ket{1}\!\bra{0}$.

When the qubit $S$ gets entangled with an environment $E$,
its state becomes mixed. This implies that another term should be included
in the previous relation, which then turns into~\cite{jakob-2003}:
\beq
 \cc_{SE}+\pp_S+\vv_S =1, \label{compl}
 \eeq
 where $\conc_{SE}$
is the concurrence~\cite{wootters}, which measures the entanglement
between $S$ and $E$. Independently of the dimension of $E$, the
bipartite concurrence of the pure composite state is defined as
\begin{equation}
\conc_{SE}=\sqrt{2(1-\tr[\rho_S^2])},
\label{eq:concdef}
\end{equation}
 where
$\rho_S=\tr_E[\rho_{SE}]$ \cite{rungta01}.

 We see from
Eq.~(\ref{compl}) that whenever the entanglement
between the two systems increases, the single-particle features are
reduced. When $\conc_{SE}=1$, the visibility and
predictability vanish -- the single-qubit state has then completely
decohered.

Relation (\ref{compl}) was tested experimentally, using nuclear
magnetic resonance techniques, in Ref.~\cite{peng:052109}. We
present in section \ref{sec:results_qubit} experimental results for the
dynamics of these three quantities, obtained with a linear optics
setup. The complementarity relations among them will help us to
understand the action  of different types of environments on qubits.

\subsection{Quantum channels}

We describe now some of the most usual channels for qubits; amplitude damping, dephasing, bit-flip, phase-flip, and bit-phase-flip.  

\subsubsection{Amplitude damping}
\label{deco1}

This channel represents the dissipative interaction between the qubit and its environment. The emblematic example is given by the spontaneous emission of a photon by a two-level atom into a zero-temperature environment of  electromagnetic-field modes.

A simple way to gain insight about this process is through the corresponding quantum map:
\bea{ccl}
\label{AmplitudeDampingMap}
\ket{0}_S\ket{0}_E&\rightarrow& \ket{0}_S\ket{0}_E\;;\\
\ket{1}_S\ket{0}_E&\rightarrow&\sqrt{1-p}\ket{1}_S\ket{0}_E + \sqrt{p}\ket{0}_S\ket{1}_E\;,
\eea
which can be traced back to the 1930 Weisskopf-Wigner treatment of spontaneous emission by an atom~\cite{weisskopf}.

The first line indicates that if no excitation is present in the system, it remains in the same  state and the environment is also untouched. The next line shows that when one excitation is present in the system, it can either remain there with probability $(1-p)$, or it can be transferred into the environment with a probability $p$.

Notice that $p$ in these equations is just a parameterization of
time. The relationship between the parameter $p$ and time $t$ for an
atom interacting with an infinite number of electromagnetic field
modes, initially in the vacuum state, under the Markov approximation, is given by $p=(1-e^{-\Gamma
t})$, where $\Gamma$ is the decay rate. In this case, the state
$\ket{1}_E$ in the map above can be understood as one excitation
distributed in all field modes. However, this map can also be used
to describe the interaction of a two-level atom with a single mode
of the electromagnetic field inside a high-quality
cavity~\cite{raimond:565}. In this case the excitation oscillates
between the atom and the field, and  we should take $p=\sin^2(\Omega
t/2)$, where $\Omega$ is the vacuum Rabi frequency. 

The fact that the same set of equations describes the interaction with either a reservoir or an environment with a single degree of freedom is a consequence of the general character of the Kraus approach, as commented right after Eq.~(\ref{KrausMaps}). 

These remarks show that it is actually very advantageous to describe the evolution of the system through a quantum channel, rather than through a specific master equation or Hamiltonian. Together with the  parameterization of the evolution in terms of $p$, thus avoiding a specific time dependence, this leads to a very general description, which includes many different processes in the same framework.

In all cases, the dynamics represented by the map \eqref{AmplitudeDampingMap} has the following Kraus operators
(in the computational basis $\{|0\rangle,|1\rangle\}$):

 \begin{eqnarray}
M_0=\left(\begin{array}{cc}
           1&0\\
           0&\sqrt{1-p}
           \end{array}\right) &\;& M_1=\left(\begin{array}{cc}
                                           0&\sqrt{p}\\
                                           0&0
                                          \end{array}\right).
\label{Kraus1A}
\end{eqnarray}

Let  $\ket{\chi}=\alpha\ket{0}+\beta\ket{1}$ be a general initial qubit state, i.e, at $p=0$. According to Eq.~(\ref{EvolKraus}), it evolves under the amplitude channel to:
\beq
\$(\ket{\chi}\!\bra{\chi})=\left(\begin{array}{cc}
|\alpha|^2+p|\beta|^2&\alpha\beta^*\sqrt{1-p}\\
\alpha^*\beta\sqrt{1-p}&(1-p)|\beta|^2
\end{array}\right).
\eeq
We can see from this state that coherence decreases with increasing $p$. Also, the population of $\ket{1}$  is transferred to $\ket{0}$. When describing the spontaneous decay ($p=1-e^{-\Gamma t}$), only in the asymptotic limit $t\rightarrow \infty$ coherence drops to zero and the system tends to the ground state. These conclusions can also be drawn from the expressions for the visibility and the predictability:
\bea{ccl}
{\cal P}_S(p)& =& ||\alpha|^2-|\beta|^2+2 p |\beta|^2|=|1-2 (1-p) |\beta|^2|\;;\\
{\cal V}_S(p) &=& 2 \sqrt{1-p}|\alpha\beta|=\sqrt{1-p}\;{\cal V}_S(0)\,,
\label{eq:predvis}
\eea
where ${\cal V}_S(0)$ is the initial visibility. Furthermore, within the entire interval $0<p<1$ the qubit state is mixed.  This is confirmed by the calculation of its entanglement with the environment:
\beq
\conc_{SE}(p)=2|\beta|^2\sqrt{p(1-p)}\;;
\label{eq:concevol}
\eeq
which vanishes only at $p=0$ or $p=1$.\\

\subsubsection{Dephasing}\label{deco2}
{
Here the coherence of the qubit state disappears without any  change in the populations.  This process occurs often when a noisy field couples to a two-level system~\cite{leibfried:281}.

The corresponding unitary evolution map is given by:
\bea{ccl}
\ket{0}_S\ket{0}_E&\rightarrow& \ket{0}_S\ket{0}_E\;,\\
\ket{1}_S\ket{0}_E&\rightarrow&\sqrt{1-p}\ket{1}_S\ket{0}_E + \sqrt{p}\ket{1}_S\ket{1}_E.
\label{PhaseDampingMap}
\eea
It can be understood as an elastic scattering, where the the state of the two-level system does not change, but the state of the environment undergoes a transition without any energy exchange, due for instance to the change of momentum of its constituent particles. Although the states of the computational basis $\{|0\rangle,|1\rangle\}$ do not change under this map, any superposition of them will get entangled with the environment.

The characteristics of this type of channel can be analyzed, as
before, by observing  the evolution of a general state. The corresponding Kraus
operators are:
\begin{eqnarray}
M_0=\left(\begin{array}{cc}
           1&0\\
           0&\sqrt{1-p}
           \end{array}\right) &\;& M_1=\left(\begin{array}{cc}
                                           0&0\\
                                           0&\sqrt{p}
                                          \end{array}\right).
\end{eqnarray}
Therefore, the state $\ket{\chi}=\alpha\ket{0}+\beta\ket{1}$ evolves to:
\beq
\left(\begin{array}{cc}
|\alpha|^2&\alpha\beta^*\sqrt{1-p}\\
\alpha^*\beta\sqrt{1-p}&|\beta|^2
\end{array}\right).
\eeq 

As previously stated, the populations do not change,  as well
as the state predictability:
$\prev_S(p)=||\alpha|^2-|\beta|^2|=\prev_S(0)$. On the other hand,
the visibility monotonically decreases:  $\visi_S(p) =
2|\alpha\beta|\sqrt{1-p}=\sqrt{1-p}\,\visi_S(0)$, as the system $S$
gets entangled with the environment $R$. The entanglement between them
is easily evaluated: $\conc_{SE}(p) = 2 \sqrt{p}|\alpha\beta| =
\sqrt{p}\,\visi_S(0)$. This emphasizes the fact that states with
zero initial visibility do not get entangled with this type of
environment.
 \begin{table*}
 \caption{\label{ErrorChannels} Evolution of  complementary aspects for the initial state $\ket{\chi}=\alpha\ket{0}+\beta\ket{1}$ under bit, phase, and bit-phase flip.}
 \begin{ruledtabular}
 \begin{tabular}{c|ccc}
Channel  & $\prev_S(p)$ & $\visi_S(p)$ & $\conc_{SR}(p)$  \\
\hline
Bit flip & $(1-p)\;{\cal P}_S(0)$ &$|(2-p)\alpha\beta^*+p \alpha^*\beta|$ & $ \sqrt{p\left( 2-p \right)}|\alpha^2-\beta^2| $ \\
Phase flip & ${\cal P}_S(0)$ &$(1-p){\cal V}_S(0)$ & $\sqrt{p(2-p)}{\cal V}_S(0)$  \\
Bit-Phase flip & $(1-p){\cal P}_S(0)$
&$|(2-p)\alpha\beta^*-p\;\alpha^*\beta|$ &
$\sqrt{p(2-p)}|\alpha^2+\beta^2|$
\end{tabular}
 \end{ruledtabular}
 \end{table*}

\subsubsection{Bit flip, phase flip, and bit-phase-flip}

In classical computation,  the only error that can take place is the bit flip $0\leftrightarrow 1$. In quantum computation however, the possibility of superposition brings also the possibility of other errors besides the usual bit flip. They are the phase flip and the bit-phase flip. The first changes the phase of the state, and the latter combines phase- and bit-flip.

The set of  Kraus operators for each one of  these channels is given by:
 \begin{eqnarray}
M_0= \sqrt{1-p/2}\;\openone\,, &\;& M_1^i=\sqrt{p/2}\;\sigma_i\;;
\label{KrausErrors}
\end{eqnarray}
where $i=x$ give us the bit flip, $i=z$ the phase flip, and $i=y$ the phase-bit flip. These sets are easily interpreted as corresponding to a probability $(1-p/2)$ of remaining in the same state, and a probability $p/2$ of having  an error. The factor of 2 in Eq.~(\ref{KrausErrors}) guarantees that at $p=1$ we have maximal ignorance about the occurrence of an error, and therefore minimum information about the state. The unitary maps for these channels are obtained by employing Eq.~(\ref{KrausMaps}).

In Table~\ref{ErrorChannels}, the evolution of the complementary aspects, as previously defined, are summarized for these error channels.

\subsection{Entanglement dynamics}
\label{sec:EntanglementDynamicsTheory}

Whenever the system $S$ is composed of at least two subsystems,  an initial entanglement among the subsystems evolves due to the interaction with the environment~\cite{karol,simon, diosi, dodd, dur, yu1, carvalho:230501, hein, fine:153105, mintert,  santos:040305, yu:140403, almeida07, carvalho-2007, kimble07, aolita}. The detailed study of this process is of crucial importance for the implementation of quantum algorithms that rely on  entanglement as a resource.

Here we focus on two emblematic examples of entanglement evolution: the two qubit state $\ket{\phi}=\alpha \ket{00}+\beta\ket{11}$ under local {\it i)} amplitude damping, and {\it ii)} dephasing channels.   In the following analysis, the complementarity relation is not easily handled~\cite{tessier:107,peng:052109}, since it involves multipartite entanglement of mixed states. Nevertheless, in order to scrutinize the  dynamics, we make use of similar figures of merit, namely: the bipartite visibility ($\visi_{S_1S_2}$), the concurrence between the subsystems ($\conc_{S_1S_2}$), and the concurrence ($\conc_{SE}$) between $S=S_1\otimes S_2$ and $E=E_1\otimes E_2$. The definitions of these quantities follows.

The bipartite visibility,
\beq
\visi_{S_1S_2}(p)=2 |\langle\, \ket{11}\!\bra{00} \, \rangle|\;,
\label{eq:bivis}
\eeq
 measures the two-particle coherence for the state $|\phi\rangle$ defined above. Notice that, given the initial state $\ket{\phi}$, and the fact that we are considering only local channels, this is the only coherence that plays a role in the dynamics.
 
The initial pure state of  the system $S=S_1\otimes S_2$ becomes mixed when in contact with the environment. The degradation of the initial entanglement due to the coupling with the environment is quantified by the concurrence defined in Ref.~\cite{wootters}:
 \begin{equation}\label{concurrence}
\conc_{S_1S_2}(p) =\max \{0,\Lambda\}\,,
\end{equation}
where  $ \Lambda=\sqrt{\lambda_1}-\sqrt{\lambda_2}-\sqrt{\lambda_3}-\sqrt{\lambda_4}$, with $\lambda_i$'s the eigenvalues in decreasing order of:
\begin{equation}
\rho_{S_1S_2}(p)(\sigma_y\otimes\sigma_y)\rho^*_{S_1S_2}(p)(\sigma_y\otimes\sigma_y)\,,
\end{equation}
 the conjugation being taken in the computational basis $\{|00\rangle, |01\rangle, |10\rangle, |11\rangle\}$, and  $ \rho_{S_1S_2}(p) = \$_1\otimes \$_2 (\ket{\phi}\!\bra{\phi})$, where $\$_1$($\$_2$)  is the channel applied to the first (second) qubit.

The information spread from the initial pure state to the combined state -- system plus environment -- is related to the entanglement between $S$ and $E$.  The corresponding concurrence is~\cite{rungta01}:
\beq
\conc_{SE}(p)=\sqrt{2\left(1-\tr\left [\rho^2_{S_1S_2}(p)\right]\right)}\;.
\eeq

 {\it i) Amplitude damping --} As described before in
section~\ref{deco1}, the Kraus  operators for this channel are given
in Eq.~(\ref{Kraus1A}). Under two identical local amplitude channels, Eq.~(\ref{EvolKrausN}) shows that the initial two-qubit state $|\phi\rangle$ evolves to the density operator
\small \beq \left(\begin{array}{cccc}
|\alpha|^2+p^2|\beta|^2&0&0&(1-p)\alpha\beta^*\\
0&(1-p)p|\beta|^2&0&0\\
0&0&(1-p)p|\beta|^2&0\\
(1-p)\alpha^*\beta&0&0&(1-p)^2|\beta|^2
\end{array}\right)\;,
\label{evolS1S2A}
\eeq
\normalsize
where the matrix is written in terms of the computational basis $\{|00\rangle, |01\rangle, |10\rangle, |11\rangle\}$.

The two-particle visibility is then $\visi_{S_1S_2}(p)=2 (1-p)|\alpha\beta|=(1-p)\visi_{S_1S_2}(0)$. The bipartite visibility decays linearly with $p$, reaching zero only when $p=1$.

For the entanglement between the subsystems, we have:
\beq
\conc_{S_1S_2}(p)=\max\{0,2(1-p)|\beta|(|\alpha|-p|\beta|)\}\,.
\label{concESD}
\eeq
For   the same initial concurrence ($\conc_{S_1S_2}(0)=2|\alpha\beta|$), two entanglement decay regimes are found: if $|\alpha|\ge |\beta|$,  then $\conc_{S_1S_2}(p)>0$ for all $p\in[0,1)$, vanishing only at $p=1$ (as the visibility). However, for $|\alpha| < |\beta|$ the entanglement between $S_1$ and $S_2$ goes to zero at $p_{ESD}=|\alpha/\beta|$ -- the so-called \emph{entanglement sudden-death}~\cite{yu:140403, almeida07}. If the parameterization $(1-p)=e^{-\Gamma t}$ is used, this implies a finite-time disentanglement, even though the bipartite coherence goes to zero only asymptotically. This phenomenon stresses that bipartite coherence is necessary for entanglement but does not coincide with it -- the latter being more fragile  to noise.
\begin{figure}[t]
\centering
\resizebox{!}{4.8cm}{\includegraphics{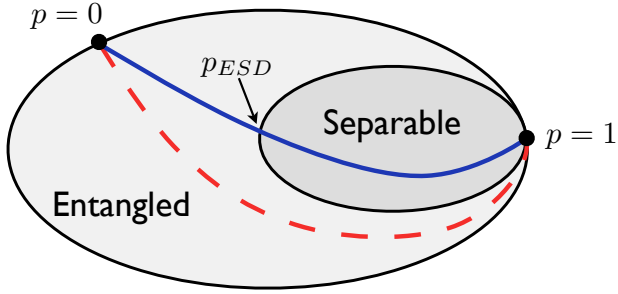}}
\caption{Two possible trajectories in the space of states under the action of amplitude damping, for initial states of the form $\alpha|00\rangle+\beta|11\rangle$. The solid line represents a sudden-death trajectory, and the dashed line a case of infinite-time disentanglement. When $p=1$, the two qubits are in the ground state. The border of the set of states is the locus of density matrices of incomplete rank.}
\label{fig:TrajAmp}
\end{figure}

Entanglement sudden death requires the initial population of the doubly excited state $|11\rangle$ to be larger than the population of the unexcited state $|00\rangle$. This is related to the fact that the state $|11\rangle$ is perturbed by the (zero temperature) environment, while the state $|00\rangle$ is not. Therefore,  the bigger the initial  ``excited" component in $\ket{\phi}$, the stronger  is the entanglement with the environment -- thus  leading to a faster decay of  $\conc_{S_1S_2}$. Indeed, the entanglement between the system and the environment is given by:
 \beq
\conc_{SE}(p)=2\sqrt{2}|\beta|\sqrt{p(1-p)}\sqrt{1-|\beta|^2 p
(1-p)},
\eeq
which increases when $\beta$ increases and, for fixed $\beta$, reaches its maximum for $p=1/2$.  This behavior is further stressed by realizing that the entanglement between each system and its own environment is also proportional to the excited-state amplitude:\\
\beq
\conc_{S_1E_1}(p)=\conc_{S_2E_2}(p)= 2 |\beta|^2 \sqrt{p(1-p)}\;;
\eeq
vanishing only at $p=0$ and $p=1$.

These two possible ``trajectories"~\cite{terra:237,yu:2289} in the set of states are sketched in Fig.~\ref{fig:TrajAmp}. For $|\alpha|<|\beta|$ (solid line) the set of separable states is crossed at $p_{ESD}$, thus the state becomes separable at finite time. However, for  $|\alpha| \ge  |\beta|$ (dashed line),  the state becomes separable only at $p=1$, when the two qubits are in the ground state ($\ket{00}$). 

This type of environment acts as a swapping process at $p=1$, i.e., the state of the system (and the corresponding entanglement) is completely transferred to the environment~\cite{yonac:s45,lopez-2008}:
\begin{equation}
\left(\alpha\ket{00}+\beta\ket{11}\right)_S\otimes\ket{00}_R \stackrel{p=1}{\longrightarrow}
\ket{00}_S\otimes\left(\alpha\ket{00}+\beta\ket{11}\right)_R\;.
\eeq
The entanglement between the two environments is given by:
\beq
\conc_{E_1E_2}(p)=\max\{0,2p|\beta|(|\alpha|-(1-p)|\beta|)\},
\eeq
which shows that whenever there is entanglement sudden death for the two-qubit system,  there is also sudden birth of entanglement (ESB) between the two corresponding environments~\cite{lopez-2008}. The value of $p$ for which ESB occurs, $p_{ESB}$, is simply expressed in terms of the entanglement sudden death value:  $p_{ESB}=1-p_{ESD}$. This expression clearly shows that entanglement sudden birth may occur before, simultaneously, or after entanglement sudden death, depending on whether $p_{ESD}>1/2$, $p_{ESD}=1/2$, or $p_{ESD}<1/2$, respectively.

{\it ii) Dephasing --} The evolved state in this case is given by:
\beq
\rho(p)=\left(\begin{array}{cccc}
|\alpha|^2&0&0&(1-p)\alpha\beta^*\\
0&0&0&0\\
0&0&0&0\\
(1-p)\alpha^*\beta&0&0&|\beta|^2
\end{array}\right).
\label{evolS1S2Dp}
\eeq

As above, the two-particle visibility is given by  $\visi_{S_1S_2}(p)=2 (1-p)|\alpha\beta|=(1-p)\visi_{S_1S_2}(0)$, which leads to the same behavior as before.

The entanglement between the subsystems is:
\beq
\conc_{S_1S_2}(p)=2 (1-p)|\alpha\beta|\;,
\eeq
which is precisely equal to $\visi_{S_1S_2}(p)$. These two quantities have the same behavior as a function of $p$, and  thus vanish at same point $p=1$. There is no entanglement sudden-death.

The entanglement between the system $S$ and the environment $E$ is given by:
\beq
\conc_{SE}(p)=2|\alpha\beta|\sqrt{p(2-p)}\;;
\eeq
which reaches its maximum, for $p$ fixed, when $|\alpha|=|\beta|=1/\sqrt{2}$. For every $\alpha$ and $\beta$ the maximum of  $\conc_{SE}$ as a function of $p$ is at $p=1$, i.e, only when dephasing is completed. However, increasing values of $\conc_{SE}$ do not imply sudden death of entanglement, since the corresponding state trajectory does not cross the region of separable states (see Fig.~\ref{fig:TrajDep}).

\begin{figure}[t]
\centering
\resizebox{!}{5.2cm}{\includegraphics{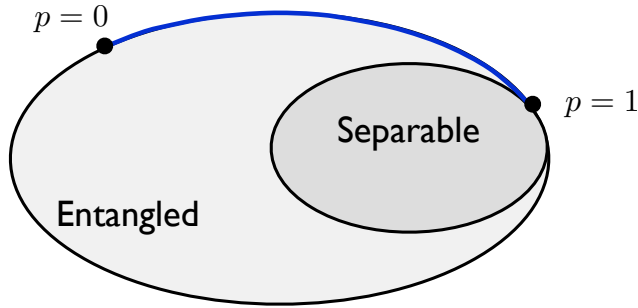}}
\caption{Trajectory  in the space of states under the action of dephasing, for initial states of the form $\alpha|00\rangle+\beta|11\rangle$. The state is completely decohered at $p=1$, when it reaches the borderline between entangled and separable states -- and only then becomes separable. For this case, the trajectory stays always on the border of the set of states, since, for all $p\in[0,1]$, the density matrix is not of complete rank.}
\label{fig:TrajDep}
\end{figure}

One should note that, in contrast with the amplitude-damping case, here each system does \emph{not} get entangled with its own environment: $C_{S_1E_1}=C_{S_2E_2} = 0$ for all $p \in [0,1]$.  This is expected, since the single-qubit visibility is zero at the beginning, and consequentially at all subsequent times (see Section~\ref{deco2}). What is though more surprising is that, apart from $\conc_{S_1S_2}$ which monotonously decreases with $p$, all other two-quibt entanglements are identically zero. Decrease in the entanglement of the two-qubit system is accompanied by the creation of legitimate multipartite entanglement. 

For $p=1$, it is easy to see from Eq.~(\ref{PhaseDampingMap}) that 
\beq
(\alpha|00\rangle+\beta|11\rangle)_S|00\rangle_{E}
\rightarrow\alpha|00\rangle_S|00\rangle_E+\beta|11\rangle_S|11\rangle_E\,,
\eeq
which is a Greenberger-Horne-Zeilinger (GHZ) type~\cite{GHZ} type of state, for which any two-qubit entanglement vanishes. 

For arbitrary values of $p$, one can easily calculate the generalized multipartite concurrence proposed in Ref.~\cite{mintert}:
\begin{equation}
{\cal C}_{N}=2^{1-N/2} \sqrt{(2^{N}-2)-\sum_{i} {\rm Tr}(\rho_{i}^{2})}
\end{equation}
where the sum is over all nontrivial reduced density matrices of the N-particle system. We get 
\beq
\conc_{S_1S_2E_1E_2}(p)= |\alpha\beta|\sqrt{4+4 p -p^2}\;
\eeq
which monotonously increases with $p$.

%
%
\section{Experimental Implementation of Decoherence Channels}
\label{sec:experiment}

\subsection{Single photons and multiple qubits}
Many experimental investigations of quantum information processes, such as  basic quantum algorithms \cite{cerf98,oliveira05,walborn05c}, quantum teleportation \cite{boschi98}, and verification of new methods for measuring entanglement~\cite{walborn06b,walborn07a} have been based on the use of several degrees of freedom of  single photons.  While this type of approach does not lead to scalable quantum computation \cite{blume-kohout02}, taking advantage of multiple degrees of freedom of photons allows for entanglement purification \cite{pan03}, improved Bell-state analysis \cite{kwiat98a,walborn03b,walborn03c,kwiat07} and creation of high-dimensional entanglement \cite{barreiro05}. The extra degrees of freedom have also been exploited to engineer mixed states through decoherence \cite{peters04,aiello07}.
\par
In the following we employ the polarization degree of
freedom of a photon as the qubit, while its  momentum degree of
freedom is used as the environment.   This  choice enables us to implement controlled interactions between $S$ and $E$. As in previous works~\cite{cerf98,oliveira05}, it is
possible to implement a variety of operations on these two degrees of
freedom, using common optical elements such as wave plates and beam
splitters.  The formal correspondence between linear optics
operations and one or multiple-qubit quantum operations has been
provided in Ref. \cite{aiello07b}.
\par
\subsection{Sagnac Interferometer}

Fig. \ref{fig:int} a) shows a modified Sagnac interferometer  that can
be used to implement the dynamics discussed in section
\ref{sec:theory}. An incident photon passes through a polarizing
beam splitter (PBS), which splits the horizontal ($H$) and vertical
($V$) polarization components, causing them to propagate in opposite
directions within the interferometer.  The interferometer is aligned so
that the $H$-path and $V$-path are slightly separated, which allows
us to insert different optical elements separately into each path.
The two paths then recombine at the same PBS, and are reflected or
transmitted into modes $0$ or $1$, depending on the polarization. 
HWP($\theta_{H}$) and HWP($\theta_{V}$)  rotate  the $H$ and $V$
polarization components of the incoming photon, respectively. If they are set at
positions such that the polarizations are not rotated, the photon
leaves the Sagnac interferometer in mode $0$.   If, however, a
photon, initially $V$-polarized, is rotated by HWP($\theta_{V}$) so that $\ket{V}
\longrightarrow \alpha \ket{V} + \beta\ket{H}$, it will leave the
interferometer in mode $0$ with probability $|\alpha|^2$, and in
mode $1$ with probability $|\beta|^2$. 
The Sagnac arrangement is  advantageous, since it is very robust
against small mechanical fluctuations of the mirrors and polarizing
beam splitter  (photons in the two paths reflect off the same optical
components), as well as thermal fluctuations. The two optical paths are approximately the same, since they have identical lengths and both include a single half-wave plate. We  now discuss the implementation of decoherence channels with this interferometer.
  \begin{figure}[t]
 {\includegraphics[width=7cm]{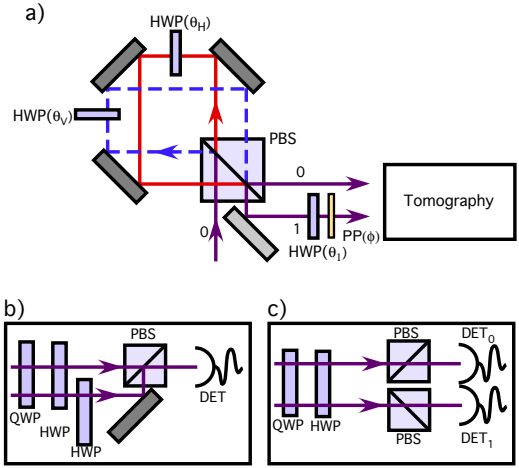}}%
 \caption{ Experimental apparatus for implementation of quantum maps and tomographic analysis.  a) Sagnac interferometer. HWP($\theta_{V}$), HWP($\theta_{H}$) and HWP($\theta_{1}$) are half-wave plates aligned in such a way that  $\theta_{V}$ ($\theta_{H}$, $\theta_{1}$) is measured with respect to V-polarization (H-polarization), PBS is a polarizing beam splitter, PP$(\phi)$ is a phase plate and unlabelled rectangles are mirrors. Polarization tomography is then performed on the output modes. b)  Tomography setup used when tracing out the environment modes.  The quarter- and half-wave plates QWP and HWP are used for the tomographic process and HWP$(45^\circ)$ is a half-wave plate used to incoherently recombine the environment modes at the PBS.  c) Tomography setup used when monitoring the environment.  Detectors DET$_0$ and DET$_1$ detect photons in modes $0$ and $1$, respectively.}
 \label{fig:int}
 \end{figure}

\subsubsection{Decoherence channels}

With the half-wave plates set to angles $\theta_H$ and $\theta_V$, the Sagnac interferometer implements the transformation: 
\begin{subequations}
\label{eq:ampdampexp1}
\begin{align}
\ket{H}\ket{0} & \longrightarrow \cos2\theta_H\ket{H}\ket{0} +\sin2\theta_H\ket{V}\ket{1} \,,\\
\ket{V}\ket{0} & \longrightarrow \cos2\theta_V \ket{V}\ket{0} + \sin
2\theta_V\ket{H}\ket{1}.\label{ampdampb}
\end{align}
\end{subequations}
After the  half-wave plate and the phase plate in output mode $1$, the overall transformation is
\begin{subequations}
\label{eq:ampdampexp2}
\begin{align}
\ket{H}\ket{0}  \longrightarrow & \cos2\theta_H\ket{H}\ket{0} +e^{i\phi}\sin2\theta_H\sin2\theta_1\ket{H}\ket{1} \nonumber \\
& -e^{i\phi}\sin2\theta_H\cos2\theta_1\ket{V}\ket{1} \\
\ket{V}\ket{0}  \longrightarrow & \cos2\theta_V \ket{V}\ket{0} + e^{i\phi}\sin 2\theta_V\cos2\theta_1\ket{H}\ket{1} \nonumber \\
& + e^{i\phi}\sin 2 \theta_V\sin2\theta_1\ket{V}\ket{1}.
\end{align}
\end{subequations}

By associating $H$ and $V$ polarizations respectively to the ground and excited states of the qubit, output modes $0$ and $1$ to states of the environment, and
adequately choosing the correct wave plate angles, a number of decoherence channels can be implemented with this interferometer.  For example, setting $\theta_H=0$, $\theta_1=0$, $\phi=0$ and identifying $p=\sin^2 2\theta_V$, the interferometer corresponds to the amplitude damping channel \eqref{AmplitudeDampingMap}.   

Using the same settings but with $\theta_1=\pi/4$ implements the phase damping channel \eqref{PhaseDampingMap}.  Also, the error channels shown in Table \ref{ErrorChannels} can be implemented.  For example, $\theta_H=-\theta$, $\theta_V=\theta$, $\theta_{1}=\phi=0$  implements a bit-flip channel with $p=\sin^2 2\theta_V$.  Table \ref{tab:1}  shows the wave plate settings for several different decoherence channels.
\par

\label{sec:decoherencechannels}
 \begin{table}
 \caption{\label{tab:1} Wave plate angles and phase $\phi$ for different decoherence channels.}
 \begin{ruledtabular}
 \begin{tabular}{ccccc}
Channel  & $\theta_H$ & $\theta_V$ & $\theta_1$ & $\phi$ \\
\hline
Amplitude decay & 0 & $\theta$ & 0&  0 \\
Phase decay & 0 &$\theta$ & $\pi/4$ &  0 \\
Bit flip & $-\theta$ &$\theta$ & 0&  0 \\
Phase flip & $\theta$ &$-\theta$ & $\pi/4$ &  0 \\
Bit-Phase flip & $-\theta$ &$-\theta$ & $0$ &  $\pi/2$
\end{tabular}
 \end{ruledtabular}
 \end{table}

In order to investigate decoherence using this interferometer, it is necessary to combine modes $0$ and $1$ incoherently for detection, which is the experimental equivalent to mathematically ``tracing out" the environment.  This was done using the two input ports of the PBS used for polarization tomography, as shown in Fig. \ref{fig:int} b).  Incoherent combination is guaranteed as long as the path length difference is larger than the coherence length of the photon.
Using two or more interferometers, one can use similar setups to study the evolution of multipartite states subject to different combinations of independent error channels.

\subsubsection{Monitoring the environment}
\label{sec:environmentMonitoringSetup}

As opposed to uncontrolled physical processes that induce decoherence in other
systems, our interferometric arrangement allows us to monitor the
environment. This can be done for instance by simply detecting photons in momentum modes $0$ or $1$ individually, as illustrated in Fig. \ref{fig:int} c).   With this setup we are able to experimentally investigate the filtering operations and the quantum jumps described in section \ref{sec:environmentMonitoringTheory}.  The corresponding experimental results are discussed in the next section.

\section{Experimental Results:  Single-Qubit Decay and the dynamics of complementarity relations}
\label{sec:results_qubit}
In  the experiments reported in this section and the next, we controlled the system-environment interaction by varying  the parameter $p$.  For each value, we performed full tomography of the single or two-photon polarization state and reconstructed the density matrix using the maximum likelihood method  \cite{kwiat-tomo,altepeter}.  The purity and the concurrence were obtained from the reconstructed density matrix unless otherwise noted.  The theoretical predictions were obtained by evolving the reconstructed initial state, corresponding to $p=0$, using the Kraus operator formalism discussed in section \ref{sec:theory}.   Vertical experimental error bars were determined by Monte-Carlo simulation of experimental runs obeying the same Poissonian count statistics.  The value of $p$ was determined by one of two methods.  In our earlier experiments, we used the direct readout of the angle (larger horizontal error bars, due to coarse angular setting).  In later experiments, we improved on this by developing a simple way to determine $p$ empirically, which we now describe. First, we block the interferometer arm corresponding to the $H$ component [propagation in the counterclockwise direction in Fig. \ref{fig:int} a)], and measure the counts $c_0$ in output mode $0$,
with the tomographic plates set for measuring $V$ polarization.  Then, still blocking the $H$ interferometer
arm, we measure the counts $c_1$ in output mode $1$, with the tomography
system set to $H$.  We
obtain  $p$  from $p=
{c_1}/(c_0+c_1)$. This method is more precise, since the uncertainty in $p$ comes from
photon count statistics.  
\par
\subsection{Amplitude damping channel} 
For the study of the amplitude decay of a single qubit, we use a c.w. solid state laser (405 nm)
to pump a 5 mm long LiIO$_3$ non linear crystal, producing photon pairs from spontaneous parametric down conversion (SPDC).
The signal and idler photons are prepared in polarization product states
with $V$ polarization. Here the idler photon is used only as a trigger, and is sent directly
to a detector equipped with an interference filter centered at 800nm, (65nm FWHM) and a 0.5mm diameter pinhole.  
The signal photon goes through the interferometer shown in
Fig.~\ref{fig:int} a), with wave plates aligned for implementation of the amplitude damping channel, as discussed in section \ref{sec:decoherencechannels}.  After the interferometer, modes 0 and 1 propagate through wave plates and a polarizing beam splitter, necessary in the tomography process. Afterwards they are detected through an interference filter centered around 800nm,
with 10nm  bandwidth and a 0.5mm pinhole.  
Coincidence counts are registered with counting electronics and a computer.    
 
\par

The amplitude damping channel was implemented for a single qubit, with the detection system set to trace over the modes of the environment, using the detection setup shown in Fig.~\ref{fig:int} b).  The input polarization was prepared in a superposition state $\alpha\ket{H} + \beta \ket{V}$, with $|\beta|>|\alpha|$.  
\par
It is  illustrative to view the effect of the channel by measuring the quantities involved in the complementarity relation discussed
in section~\ref{sec:complementarity}. In
Fig.~\ref{fig:SingleParticleAD} we show the evolution of the squared predictability ${\cal P}_S^2$, the squared visibility
${\cal V}_S^2$, and the system-environment entanglement, quantified by
the squared concurrence, ${\cal C}_{SE}^2$, as a
function of $p$ for the same initial state as
above.  The concurrence ${\cal C}_{SE}$ was calculated from the density matrix using \eqref{eq:concdef}, and coincides with what is expected from Eq. \eqref{eq:concevol}.  ${\cal P}_S$ and 
${\cal V}_S$ were determined directly from the polarization measurements using
\beq
{\cal P}_S = \frac{|c_H-c_V|}{c_H+c_V}
\eeq
\beq
{\cal V}_S = 2 \sqrt{\left(\frac{2c_+}{c_H+c_V} -1\right)^2+\left(\frac{2c_R}{c_H+c_V}-1\right)^2}
\eeq
for each value of $p$, where $c_j$ are the number of counts with $j$ polarization, with $+$ and $R$ corresponding to $45^0$ linear polarization and right circular polarization, respectively.  It can be seen from Fig.~\ref{fig:SingleParticleAD} that both quantities agree with Eqs. \eqref{eq:predvis}.  Though ${\cal P}_S^2$,
${\cal V}_S^2$, and ${\cal C}_{SE}^2$ evolve with $p$, the sum of these three quantities (yellow triangles in Fig.~\ref{fig:SingleParticleAD}) satisfies relation~\eqref{compl} for all $p$.  
\begin{figure}[t]
\centering
\resizebox{!}{5.5cm}{\includegraphics{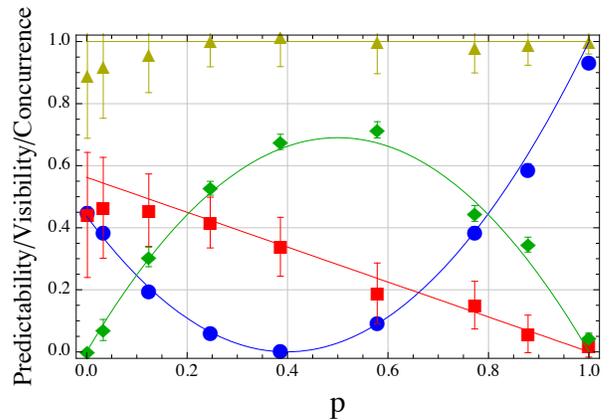}}
\caption{(Color online) Evolution of the quantities involved in the qubit-environment complementarity relation \eqref{compl} under the amplitude damping channel: ${\cal P}_S^2$ (blue circles), ${\cal V}_S^2$ (red squares) and ${\cal C}_{SE}^2$ (green diamonds).  Also shown is the sum ${\cal P}_S^2+{\cal V}_S^2+{\cal C}_{SE}^2$ (yellow triangles). The solid lines are the corresponding theoretical predictions, obtained by applying the amplitude damping channel to the experimentally reconstructed initial state.}
\label{fig:SingleParticleAD}
\end{figure}

\subsection{Monitoring the environment}
\label{sec:environmentMonitoringResults}
We demonstrate now a peculiar effect of the dynamics of open quantum systems. If the qubit, under the action of the amplitude-damping channel, is initially in a superposition of the states $|0\rangle$ and $|1\rangle$, and we monitor the state of the environment, finding it with no excitation at all times, we still observe a decay of the system towards the ground state. This can be understood as follows: even if there is no energy transfer between system and environment, by constantly monitoring the environment and finding no excitations in it, we gain information about the system, which is expressed as a change in its state.   For example, consider the arrangement used to implement the amplitude damping channel  \eqref{AmplitudeDampingMap}, with
an initial state $(\alpha\ket{H}+\beta\ket{V})_S\otimes\ket{0}_E$.  This state evolves to
\begin{equation}
|\Psi(p)\rangle=\alpha\ket{H}\ket{0} +
\beta\sqrt{1-p}\ket{V}\ket{0}+\beta\sqrt{p}\ket{H}\ket{1}.
\end{equation}
Tracing out the environment, the polarization state is
\begin{equation}
\varrho(p) = \left (\begin{array}{cc}
|\alpha|^2+|\beta|^2p & \alpha\beta^*\sqrt{1-p} \\
\alpha^*\beta\sqrt{1-p} & |\beta|^2(1-p)
\end{array}
\right )\,,
\label{eq:rhonomon}
\end{equation}
with $p^2=\sin 2\theta_V$, whereas, projecting onto the ``unexcited" $\ket{0}$ state of the environment, the polarization state becomes
\begin{equation}
\ket{\psi(p)} = \frac{\alpha\ket{H} +
\beta\sqrt{1-p}\ket{V}}{[|\alpha|^2+|\beta|^2(1-p)]^{1/2}},
\end{equation}
\par
which ``decays" to $|\psi(p=1)\rangle=|H\rangle$, just as $\varrho(p=1)$ given in \eqref{eq:rhonomon}.  We illustrate this phenomenon by comparing the dynamics of a qubit under the action of the amplitude damping map~\eqref{AmplitudeDampingMap}, for two cases: {\it (i)} when we trace out the environment's degrees of freedom (using the tomographic setup illustrated in Fig.~\ref{fig:int} b), and {\it (ii)} when we monitor its state in the unexcited state, using the tomographic setup shown in Fig.~\ref{fig:int} c).
As above, the input polarization is prepared in a superposition state $\alpha\ket{H} + \beta \ket{V}$, with $|\beta|>|\alpha|$.
\par

\par
Figure~\ref{fig:MonitoringPopulations} shows the evolution of the population $V$, for both cases. We see that not only are the two dynamics different, but also that the decay takes place even if no excitations are transferred to the environment. When the environment is traced out, the linear evolution is equivalent to an exponential decay (linear in $p$), while in the case where the environment is monitored the decay is retarded.
\begin{figure}[t]
\centering
\resizebox{!}{5.5cm}{\includegraphics{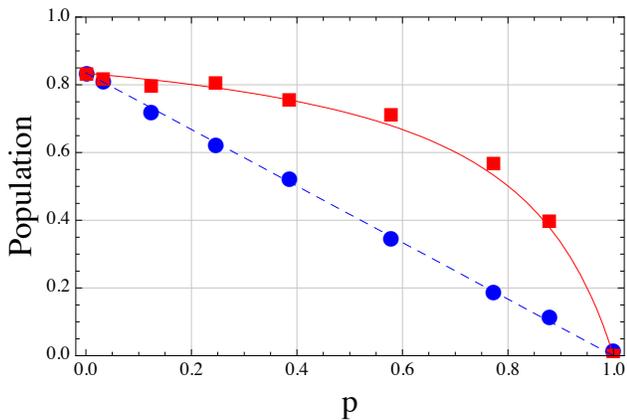}}
\caption{(Color online) Evolution of the population $V$  when monitoring (red
  squares) or tracing out (blue circles) the environment. Error bars are
  unnoticeable in this scale. The lines are
  the corresponding theoretical predictions (red solid line and blue dashed line).}
\label{fig:MonitoringPopulations}
\end{figure}

Figure~\ref{fig:MonitoringPurities} shows the evolution of the purity in these two cases.  We see that when we monitor the environment, the system is always close to a pure state. The little mixedness arises from the fact that our initial state is not perfectly pure.
\begin{figure}[t]
\centering
\resizebox{!}{5.5cm}{\includegraphics{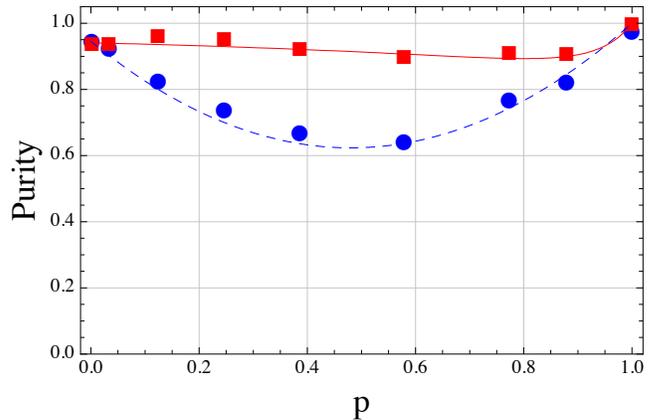}}
\caption{(Color online) Evolution of the purity of the qubit state when monitoring (red squares)
  or tracing out (blue circles) the environment. Error bars are unnoticeable in
  this scale. The lines are
  the corresponding theoretical predictions (red solid line and blue dashed line).}
\label{fig:MonitoringPurities}
\end{figure}

In these figures, the lines are the theoretical predictions, which are obtained (as introduced in section~\ref{sec:environmentMonitoringTheory}) by using
the Kraus operators corresponding to the amplitude damping channel~\eqref{AmplitudeDampingMap}. When tracing out the environment, both
operators $M_0$ and $M_1$ are used, while
when monitoring the environment in its unexcited state, only
the no-jump operator $M_0$ is used,  the resulting state
being renormalized afterwards. The agreement between theory and experimental data is quite good. These results show that, even though no excitation is transferred to the environment, the continuous acquisition of information about this fact changes the state of the qubit, increasing the probability that it is found in the unexcited state.
\par
This phenomenon allows the distillation of entanglement of a two-qubit system through continuous local monitoring of the corresponding independent environments. Indeed, for an initial state $\alpha|00\rangle+\beta|11\rangle$, with $|\alpha|<|\beta|$, continuous monitoring of the unexcited environment leads to increase of the $|00\rangle$ component, implying that the state approaches a maximally entangled state, before decaying to the state $|00\rangle$. Within the framework of the Sagnac- interferometer setup, applied to each of the two entangled photons (as described in detail in the next Section), the evolution of the system under continuous monitoring corresponds to measuring both qubits in output mode 0, while $p$ changes from $0$ to $1$. If  the two-qubit state is $|\Psi(0)\rangle=\alpha|HH\rangle+\beta|VV\rangle$ for $p=0$, then the state for $p\not=0$,  conditioned to the measurement of output mode 1 in $|0\rangle$  for both qubits, is given by:
\begin{equation}
\ket{\Psi(p)}_{\rm cond} = \frac{\alpha\ket{HH} +
\beta(1-p)\ket{VV}}{[|\alpha|^2+|\beta|^2(1-p)^2]^{1/2}}.
\end{equation}
Setting $1-p=|\alpha/\beta|$ yields a maximally entangled state. 

Therefore, continuous monitoring of the environment for the two-qubit case corresponds to a quantum distillation scheme~\cite{kwiat_distillation}.

\section{Experimental Results:  The dynamics of entanglement}
\label{sec:results_entanglement}
\begin{figure}[t]
{\includegraphics[width=8cm]{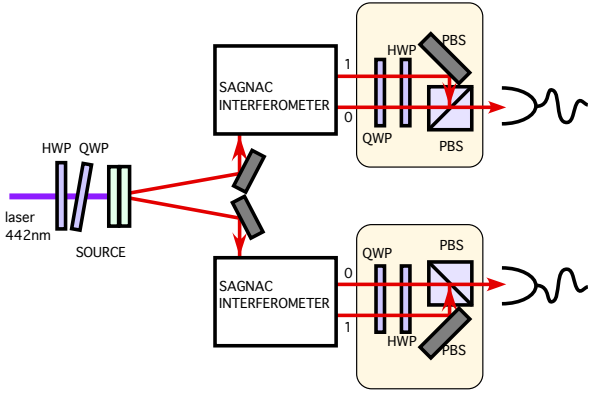}}
\caption{(Color online) Experimental setup for experimental investigation of entanglement dynamics under decoherence.  Polarization-entangled photons pairs were created using the two-crystal SPDC source \cite{kwiat99}, pumped by a 442 nm c. w. He-Cd laser. With proper spatial and spectral filtering, the 884 nm photons created in the two nonlinear crystals are prepared in an entangled state of the form $|\Theta\rangle=|\alpha||HH\rangle+|\beta|e^{i\delta}|VV\rangle$. The coefficients $|\alpha|$ and $|\beta|$, and the relative phase $\delta$ are controlled by manipulation of  the pump-beam polarization, which is easily realized with a half-wave plate (HWP) and a tilted quarter-wave plate (QWP) \cite{kwiat99}.  Tomographic analysis was performed with additional HWP's, QWP's and polarizing beam splitters (PBS).
}
\label{fig:esdsetup}
\end{figure}

Using two Sagnac interferometers, we studied the dynamical behavior of global properties of an entangled pair of photons generated with spontaneous parametric down conversion.   Fig. \ref{fig:esdsetup} shows the experimental setup.
\par
The source was arranged to generate pairs of photons in one of two non-maximally entangled states given by 
\begin{subequations}
\begin{align} 
|\Theta1\rangle& =\frac{1}{2}|HH\rangle+\frac{\sqrt{3}}{2}e^{i\theta}|VV\rangle \label{eq:state1},\\
|\Theta2\rangle& =\frac{\sqrt{3}}{2}|HH\rangle+\frac{1}{2}e^{i\theta}|VV\rangle \label{eq:state2}\,.
\end{align}
\end{subequations}

These states contain the same amount of entanglement:  the inital concurrence is ideally $C=2|\alpha\beta|=\sqrt{3}/2\simeq0.87$.  However, we measured $C=0.82\pm0.04$ and $C=0.79\pm0.11$, respectively, due to the fact that they were not 100\% pure.  The decrease in purity is mostly due to small imperfections in the mode matching of the interferometers, and the angular dependence of the phase of the two-photon state \cite{kwiat99}. To simplify the description, we refer to the initial state in the following as either (\ref{eq:state1}) or (\ref{eq:state2}), meaning in fact that it was  close to these states. The theoretical predictions are derived from the actual experimental state obtained when $p=0$, by calculating its evolution through the relevant Kraus operators.   The dynamics of entanglement was investigated under the effect of two different decoherence channels, implemented by sending the twin photons through independent Sagnac interferometers.   Full bipartite polarization-state tomography \cite{kwiat-tomo,altepeter} was performed for different values of $p$.

\subsection{Amplitude damping - Entanglement sudden death and entanglement witness}
\label{sec:GlobalAmplitudeDampingResults}
Using the wave plate configurations listed in Table \ref{tab:1}, dual  amplitude damping
maps~\eqref{AmplitudeDampingMap} with the same $p$ were implemented for the initial state (\ref{eq:state1}). 
Tomographic reconstructions of the real part of the density matrix for different values of $p$ are shown in Fig. \ref{fig:tomo1}.  The corresponding analytical expression is given by Eq.~ \eqref{evolS1S2A}.   They illustrate the evolution of the populations and coherences as a function of the parameter $p$.
   
\begin{figure}[t]
\centering
{\includegraphics[width=7cm]{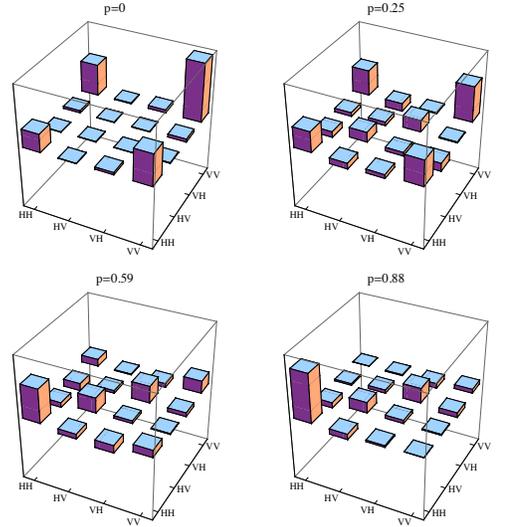}}
\caption{(Color online) Tomographic reconstruction of the real part of the density matrix for different values of $p$ for an initial state close to $\ket{\Theta1}$ given in Eq. \eqref{eq:state1}.}
\label{fig:tomo1}
\end{figure}

Figure~\ref{fig:AmplitudeDampingGlobalResults} displays the experimental results for the concurrence \eqref{concurrence}.  The theoretical prediction (denoted by the full line in the figure) was obtained by applying Eq.~ \eqref{concurrence}  to the evolved density matrix, which in turn is determined by applying the Kraus operators \eqref{Kraus1A} to the reconstructed density matrix for $p=0$.   
\begin{figure}[htbp!]
\centering
\resizebox{!}{5.5cm}{\includegraphics{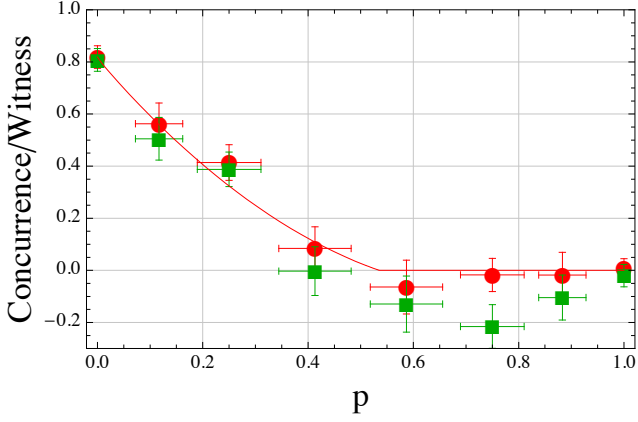}}
\caption{(Color online) Global properties under the amplitude damping channel for the state (\ref{eq:state1}):
  $\Lambda$ (as defined after Eq.~\eqref{concurrence}, red circles) and the entanglement witness \eqref{eq:Gamma} (green squares) as functions of the transition probability $p$.  The theoretical prediction for the concurrence (red solid line)
  is derived from the density matrix obtained by applying the amplitude decay channel to the experimentally reconstructed initial state. Disentanglement occurs for $p<1$.}
\label{fig:AmplitudeDampingGlobalResults}
\end{figure}
The vanishing of the entanglement for a transition probability $p<1$, corresponding to a finite time,  is clearly demonstrated in Fig.~\ref{fig:AmplitudeDampingGlobalResults}. This phenomenon was termed \emph{entanglement sudden death}~\cite{yu:140403}, and our setup allowed for its first observation, which was previously reported in~\cite{almeida07}.

\par
Also shown in Fig.~\ref{fig:AmplitudeDampingGlobalResults} are the results obtained from an entanglement witness, evaluated at each data point.
 An operator $W$ is an
entanglement witness if ${\tr}(W\rho)\ge0$ for any separable state, and
there exist entangled states $\sigma$ for which ${\tr}(W\sigma)<0$.
For initial states of the form 
\beq\label{initialstate}
|\alpha||HH\rangle+|\beta|\exp(i\theta_0)|VV\rangle
\eeq
 and the amplitude decay channel, it is possible to define a ``perfect'' $p$-independent witness~\cite{santos:040305}, so that $-{\tr}(W\rho)$ coincides precisely with $\Lambda$ in Eq.~(\ref{concurrence}), thus yielding the concurrence for all $p$. 
It is given by
\begin{equation}\label{defwitness}
\hat W_{\theta_0}   \equiv {1}- 2\left| {\Phi \left( \theta_0  \right)} \right\rangle \left\langle {\Phi \left( \theta_0  \right)} \right|\,,
\end{equation}
where
\beq\label{thetaprime}
\left| {\Phi \left( \theta  \right)} \right\rangle
= \frac{1}{\sqrt{2}}({\left| {HH} \right\rangle  + e^{i\theta} \left| {VV} \right\rangle }) \,.
\eeq
Then it is easy to show that
\beq
\Gamma_{\theta_0}\equiv -{\tr}\left[ {\hat W_{\theta_0}  \rho \left( t \right)}
\right]  = 2\left[P\left( {\theta_0 ,t}
\right)-{{1}\over{2}}\right]\,,
\label{eq:Gamma}
\eeq
where $P( {\theta ,t})=\tr[| {\Phi( \theta)}\rangle \langle {\Phi (
\theta)} |\rho(t)]$. The concurrence
\eqref{concurrence} can be written as
\[
{\cal C}\left[ {\rho \left( t \right)} \right] = {\rm{max}}\left\{ {0,\Gamma_{\theta_0} } \right\}\,.
\]
That is, the concurrence is equal to twice the excess probability (with respect to $1/2$) of projecting the system in the
maximally entangled state $\left| {\Phi \left( \theta_0  \right)}
\right\rangle$. It is remarkable that
in this case the concurrence can be given a simple physical interpretation, 
and moreover that this is valid throughout
the evolution of the system (which means, in our case, that it is independent of $p$).   The concurrence could then be determined directly 
by measuring the
probability of finding the system in the maximally-entangled
state $\left| {\Phi \left( \theta_0 \right)} \right\rangle$. 
\begin{figure}[t]
\centering
\resizebox{!}{5.5cm}{\includegraphics{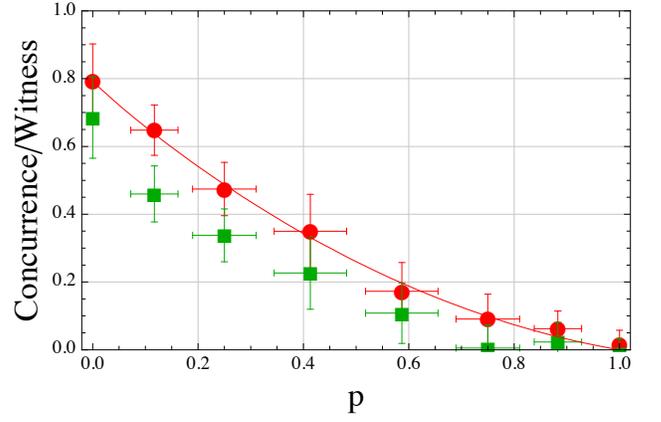}}
\caption{(Color online) Global properties under the amplitude damping channel for the state \eqref{eq:state2}: $\Lambda$ (as defined after Eq.~\eqref{concurrence}, red circles) and the entanglement witness \eqref{eq:Gamma} (green squares) as functions of the transition probability $p$.  The theoretical prediction for the concurrence (red solid line) is derived from the density matrix obtained by applying the amplitude decay channel to the experimentally reconstructed initial state. For this state, disentanglement occurs only at $p=1$. }
\label{fig:AmplitudeDampingGlobalResults2}
\end{figure}

In our experiment, however, the initial state is not pure, so $\hat W_{\theta_0}$ is not a perfect witness.  In order to compute the best witness in this case (which yields the upper bound of  $-{\tr}[\hat W_\theta\rho]$), we choose theta for each data point as the argument of the $\rho_{VVHH}$ element of the reconstructed density matrix, and then obtain $\Gamma$ through Eq.~(\ref{eq:Gamma}). The same could be achieved by projecting the state of the system on $|\Psi(\theta)\rangle$ and scanning $\theta$ in order to get the minimum value for $-{\tr}[\hat W_{\theta}\rho(t)]$. 

One should note that the initial phase $\theta_0$ is not changed by the amplitude damping channel, as  shown by Eq.~(\ref{evolS1S2A}), so in principle this procedure should be adopted only for the $p=0$ state, the corresponding witness being then valid for all values of $0<p\leq1$. However, in the experiment, changing the angle of the half-wave plate HWP$(\theta_V)$ in Fig.~\ref{fig:int} actually affects the corresponding optical path, due to imperfect alignment of this plate, so a $p$-dependent phase shows up between the states $|V\rangle|0\rangle$ and $|H\rangle|0\rangle$ on the right-hand side of Eq.~(\ref{ampdampb}). This phase does not affect the concurrence, but it implies that the best witness depends on $p$. For this reason we find  $\theta$ for each data point, from the reconstructed density matrix.  Fig. \ref{fig:AmplitudeDampingGlobalResults} shows that in this case  $\Gamma$ underestimates the entanglement.  
\par
For the state $\ket{\Theta2}$ defined in Eq. \eqref{eq:state2} the situation is drastically different.     Fig.~\ref{fig:AmplitudeDampingGlobalResults2} shows the concurrence  as  a function of $p$.  In this case the entanglement disappears only when $p=1$.  Also shown is the entanglement witness \eqref{eq:Gamma} calculated from the reconstructed density matrices, which is always less than the actual value of the concurrence.  The witness is not optimal since the initial state is not completely pure.      
\par
Figures \ref{fig:AmplitudeDampingGlobalResults} and \ref{fig:AmplitudeDampingGlobalResults2} together constitute an experimental confirmation that two states with the same initial amount of entanglement may follow different decoherence ``trajectories" in the space of states, as discussed in section \ref{sec:EntanglementDynamicsTheory} and illustrated in Fig.~\ref{fig:TrajAmp}. 

\subsection{Phase damping}
\label{sec:GlobalPhaseDampingResults}
As shown in Table \ref{tab:1}, by adjusting the wave plate angles the Sagnac interferometers implement the phase damping channel~\eqref{PhaseDampingMap}.  Experimental results for the concurrence are presented in Fig.~\ref{fig:PhaseDampingGlobalResults} for the initial state \eqref{eq:state1}.
\begin{figure}[t]
\centering
\resizebox{!}{5.5cm}{\includegraphics{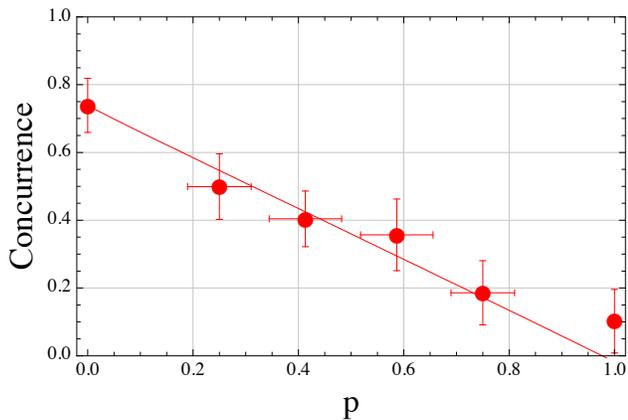}}
\caption{(Color online) Concurrence for the initial state \eqref{eq:state1} under the phase-damping channel: the circles correspond to the concurrence determined from the reconstructed density matrix for each value of $p$, while the line is the corresponding theoretical prediction, obtained by applying the phase-damping channel to the initial ($p=0$) density matrix. Disentanglement occurs only at $p=1$.}
\label{fig:PhaseDampingGlobalResults}
\end{figure}
There is no sudden death of entanglement, and concurrence vanishes only when $p=1$.

\subsection{Evolution of purity}
 
 Figure~\ref{fig:PurityEvolution}
shows the evolution of the purity of the initial state \eqref{eq:state1} for  the amplitude damping
channel~\eqref{AmplitudeDampingMap} and the phase damping
channel~\eqref{PhaseDampingMap}.  
\begin{figure}[t]
\centering
\resizebox{!}{5.5cm}{\includegraphics{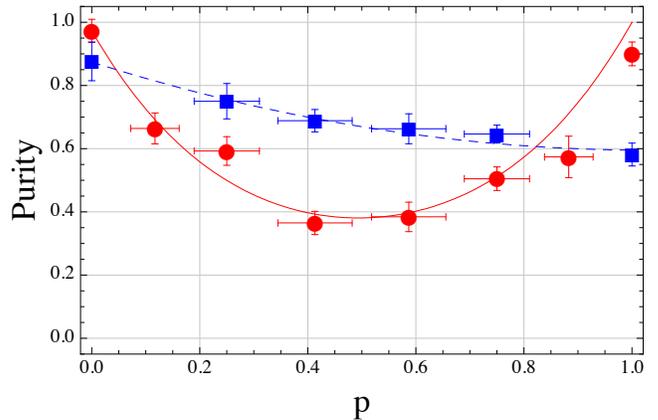}}
\caption{(Color online) Evolution of the purity for the different decoherence models: amplitude damping (red circles) and phase damping (blue squares), and their corresponding theoretical predictions.}
\label{fig:PurityEvolution}
\end{figure}
 For the phase damping channel, the change in the purity is monotonous.  For
 amplitude damping there is an initial decrease, and then it increases again up to 1 for $p=1$.  The difference in behavior  reflects the fact that amplitude damping promotes a swapping between system and environment, so that the system ends up in state $|HH\rangle$, while dephasing leads to an increase of multipartite entanglement, with the system plus environment evolving towards a GHZ-like state.

%
%

\section{Conclusions}
\label{sec:conclusions}
We have presented a series of experiments that  investigate the dynamics of entangled open quantum systems, and also the dynamics of a single qubit under continuous monitoring of the environment.  By adjusting a set of wave-plates, our linear-optics setup is capable of implementing a number of single-qubit decoherence channels.  We present experimental results for the amplitude-damping and phase-damping channels for single and two-qubit systems.  Decoherence of a single qubit is investigated through the use of complementarity relations.  The effect of decoherence on entanglement , including the phenomenon of entanglement sudden death, is experimentally demonstrated. 

Our setup has an appealing feature: it allows the investigation of filtering operations, implemented by monitoring the environment.  This is an experimental realization of quantum trajectories~\cite{carmichael}, which lead to a description of the interaction of a system with an environment in terms of pure states. For  amplitude damping, our experimental results demonstrate that it is possible to induce decay of a system by verifying, through continuous measurements, that no excitation is transferred to the environment. We have shown that this procedure, for an initially two-qubit entangled state, is equivalent to entanglement distillation. 

The experimental investigation of the environment-induced decay of entanglement in other systems  (see, for instance, Ref.~\cite{kimble07}) is of course of fundamental importance, and should help to throw new light on the subtle relation between local and global dynamics of entangled systems.

The parametrization of the quantum channels considered in this paper, in terms of the transition probability rather than time, is very convenient. It accommodates different kinds of dynamical behavior in a universal description, which allows one to extend the realm of application of the results obtained here: they include not only the decay of two-level systems interacting with individual and independent environments, but also the oscillatory exchange of energy between each qubit and another two-level system, which could be for instance the vacuum and one-photon subspace of a cavity mode.

This is a quite advantageous strategy for investigating the dynamics of disentanglement, since it pinpoints the main features of this process within a quite encompassing framework.  In fact, rather than the investigation of a particular physical system, our procedure amounts to the experimental implementation of quantum maps which, due to their generality, play a very fundamental role in quantum information.

\begin{acknowledgments}
We thank Leandro Aolita for helpful comments.
The authors acknowledge financial support from the Brazilian funding agencies CNPq, CAPES, and FAPERJ.  This work was performed as part of the Brazilian Millennium Institute for Quantum Information.  F. de M. also acknowledges the support by the Alexander von Humboldt Foundation. M. P. Almeida also acknowledges the support of the Australian Research Council and the IARPA-funded
U.S. Army Research Office Contract.
\end{acknowledgments}

%
%


\end{document}